# Aircraft trajectory control with feedback linearization for general nonlinear system

Sheng Zhang[1,2], Fei Liao[1], Yanqing Chen[1,2], Kaifeng He[1,2]
1. China Aerodynamics Research and Development Center, Mianyang, 621000
2. China State Key Laboratory of Aerodynamics, Mianyang, 621000
E-mail: zszhangshengzs@hotmail.com

**Abstract**: The feedback linearization method is further developed for the controller design on general nonlinear systems. Through the Lyapunov stability theory, the intractable nonlinear implicit algebraic control equations are effectively solved, and the asymptotically tracking performance is guaranteed. Moreover, it is proved that the controller may be used in an inverse-free version to the set-point control. With this method, a nonlinear aircraft outer-loop trajectory controller is developed. For the concern regarding the controller's robustness, the integral control technique is combined to counteract the adverse effect from modeling errors. Simulation results verify the well performance of the proposed controller.

## 1 Introduction

Feedback linearization is an important nonlinear control method [1]. This approach uses nonlinear state feedback to cancel the plant's nonlinear terms, and then linear control techniques to achieve the expected closed-loop performance. According to the research thread, generally the feedback linearization method may be divided to the differential geometry method and the dynamic inverse method [2]. The differential geometry method uses the tools including the Lie algebra, Lie bracket and differential manifold theory to realize the exact linearization, and fruitful theoretical and practical results are achieved after years of development [3-6]. The dynamic inverse method constructs the $\alpha$-order integral inverse system for the controlled plant to obtain the decoupled linear system, and it is more physically institutive for application [2][7]. Due to its effectiveness in handling nonlinear problems, the feedback linearization control method has been widely studied in aerospace engineering [8][9], electrical engineering [10][11] and mechanical engineering [12][13], etc.

One main drawback of feedback linearization control is that this method is sensitive to modeling errors and disturbances, whereas developing the exact model is usually hard in practice. Thus, the robustness issue must be considered. Various solutions are proposed such as the slide mode control [11], the adaptive control [14], the fuzzy control [15], and the observer technique [8]. Moreover, another important issue that restricts the utilization of the feedback linearization control is that it often requires the plant to be affine to obtain the analytic control law. The application to general systems is hindered for the intractable nonlinear implicit algebraic equations. This motivates us to further develop such method against the restriction. With the Lyapunov stability principle [16], the nonlinear implicit algebraic equations encountered during the controller design are addressed in a dynamics way.

Usually the aircraft guidance, i.e., the outer-loop trajectory control, and the inner-loop attitude control are designed separately based on the singular perturbation principle [17]. The guidance system realizes the commanded trajectory and sends the attitude control commands to the inner-loop attitude control system. Although commonly the aircraft trajectory control is achieved using the linear control method upon Taylor linearization [18], it is a typical nonlinear control problem in essence. To relieve the workload on the trim points selection and gain scheduling, various nonlinear control techniques are studied to enhance the performance [19][20]. In this paper, we provide another means based on the feedback linearization control. In the following, first the feedback linearization controller design for general systems is demonstrated and the tracking performance of the closed-loop control system is investigated. Then we develop the outer-loop trajectory controller for the aircraft with the proposed method. The integral control technique is also employed to increase the robustness of the feedback linearization controller. Finally simulations are carried out to examine the performance.

## 2 Feedback linearization control for general systems

### 2.1 Controller design

Consider the following nonlinear Multiple Input Multiple Output (MIMO) system

$$\dot{x} = f(x, u) \tag{1}$$

$$y = g(x) \tag{2}$$

where $x \in \mathbb{R}^n$ are the states, $u \in \mathbb{R}^m$ are the control inputs and $f$ is $n$-dimensional vector function. $y \in \mathbb{R}^m$ are the outputs and $g$ is the output function. The control objective is to track the reference signal $y_r$.

Follow the procedure in the feedback linearization controller design [1][16]. We differentiate the output equation (2) several times until the derivative expressions involve the control. Presuming that $\frac{\partial y_i}{\partial u} \neq 0$ $(i=1,2,...,m)$ is obtained for the first time after $\alpha_i$ $(i=1,2,...,m)$ times of differentiation on the output vector $y_i$ $(i=1,2,...,m)$, then we have

$$\begin{bmatrix} \frac{d^{\alpha_1} y_1}{dt^{\alpha_1}} \\ \frac{d^{\alpha_2} y_2}{dt^{\alpha_2}} \\ \vdots \\ \frac{d^{\alpha_m} y_m}{dt^{\alpha_m}} \end{bmatrix} = \begin{bmatrix} \frac{d^{\alpha_1} g_1}{dt^{\alpha_1}} \\ \frac{d^{\alpha_2} g_2}{dt^{\alpha_2}} \\ \vdots \\ \frac{d^{\alpha_m} g_m}{dt^{\alpha_m}} \end{bmatrix} \stackrel{\Delta}{=} \begin{bmatrix} v_1(x,u) \\ v_2(x,u) \\ \vdots \\ v_m(x,u) \end{bmatrix} \tag{3}$$

To achieve the desired dynamic characteristics in tracking $y_r$, we set

$$v = \begin{bmatrix} v_1 \\ v_2 \\ \vdots \\ v_m \end{bmatrix} = z \tag{4}$$

where $z$ prescribes the transition performance and it may be expressed as

$$z = \begin{bmatrix} z_1(y_1, \frac{dy_1}{dt}, ..., \frac{d^{\alpha_1-1} y_1}{dt^{\alpha_1}}, y_{1r}) \\ z_2(y_2, \frac{dy_2}{dt}, ..., \frac{d^{\alpha_2-1} y_2}{dt^{\alpha_2}}, y_{2r}) \\ \vdots \\ z_m(y_m, \frac{dy_m}{dt}, ..., \frac{d^{\alpha_m-1} y_m}{dt^{\alpha_m}}, y_{mr}) \end{bmatrix} \tag{5}$$

Presuming that the Jacobi matrix $\frac{\partial v}{\partial u}$ is full rank and the states $x$ are available by the sensors or the state observer, we then solve the following control algebraic equation to get the control law

$$v(x, u) - z = 0 \tag{6}$$

However, it is often hard to get the explicit form of $u$ for systems without affine form, and the root-finding of the implicit algebraic equation is involved to obtain the control [21][22].

Instead of employing a root-finding numerical method, here we address this issue with the Lyapunov theory, by constructing a Lyapunov function as

$$V_s = \frac{1}{2} h^T h \tag{7}$$

where $h = v - z$ and the superscript "T" denotes the transpose operator. Differentiating $V_s$ gives

$$\dot{V}_s = h^T \left( \frac{\partial h}{\partial u} \dot{u} + \frac{\partial h}{\partial x} \dot{x} + \frac{\partial h}{\partial z} \dot{z} \right) \tag{8}$$

Under the assumption that $\frac{\partial v}{\partial u}$ is full rank, if we set

$$\begin{aligned}\dot{u} &= -K_s \left( \frac{\partial h}{\partial u} \right)^T h - \left( \frac{\partial h}{\partial u} \right)^{-1} \left( \frac{\partial h}{\partial x} \dot{x} + \frac{\partial h}{\partial z} \dot{z} \right) \\ &= -K_s \left( \frac{\partial v}{\partial u} \right)^T (v - z) - \left( \frac{\partial v}{\partial u} \right)^{-1} \left( \frac{\partial v}{\partial x} \dot{x} - \dot{z} \right)\end{aligned} \tag{9}$$

where $K_c$ is a positive-definite matrix. Then we have

$$\dot{V}_s = -h^T \frac{\partial v}{\partial u} K_s \left( \frac{\partial v}{\partial u} \right)^T h \leq 0 \tag{10}$$

with the invariance principle [16], we may further proves that

$$\lim_{t \to \infty} V_s = 0 \tag{11}$$

This means that $\lim_{t \to \infty} h = 0$ and accordingly Eq. (6) will be satisfied. Thus, we may derive the explicit control law as

$$u = \int_{t_0}^{t} \left\{ -K_s \left( \frac{\partial v}{\partial u} \right)^T (v - z) - \left( \frac{\partial v}{\partial u} \right)^{-1} \left( \frac{\partial v}{\partial x} \dot{x} - \dot{z} \right) \right\} dt, \; u(t_0) = u_0 \tag{12}$$

where $u_0$ is the initial value of control at the initial time $t_0$.

For the case where no solution exists for the implicit control equation defined in Eq. (6), the pseudo-inverse [23], instead of $\left( \frac{\partial v}{\partial u} \right)^{-1}$, may be employed to adapt Eq. (12). It could be found that this treatment minimizes $V_s$ and eliminates the effect from the change of states and reference signals to the largest extent.

### 2.2 Tracking performance analysis

We will discuss the tracking performance of the proposed controller. Assuming that we know $y_r$ and its derivatives with arbitrary order, then we may obtain the error dynamics from Eq. (3) in the state-space description as

$$\dot{e} = Ae + B(v - r) \tag{13}$$

where $e = \begin{bmatrix} y_1 & \ldots & y_1^{(\alpha_1)} & y_2 & \ldots & y_2^{(\alpha_2)} & \ldots & y_m & \ldots & y_m^{(\alpha_m)} \end{bmatrix}^T - \begin{bmatrix} y_{r1} & \ldots & y_{r1}^{(\alpha_1)} & y_{r2} & \ldots & y_{r2}^{(\alpha_2)} & \ldots & y_{rm} & \ldots & y_{rm}^{(\alpha_m)} \end{bmatrix}^T$,

$r = \begin{bmatrix} y_{r1}^{(\alpha_1)} & y_{r2}^{(\alpha_2)} & \ldots & y_{rm}^{(\alpha_m)} \end{bmatrix}^T$, $A = \mathrm{diag}(A_1, A_2, \ldots A_m)$, $B = \mathrm{diag}(b_1, b_2, \ldots b_m)$, and $A_i$, $b_i$ take the form below respectively

$$A_i = \begin{bmatrix} 0 & 1 & 0 & \ldots & 0 \\ 0 & 0 & 1 & \ldots & 0 \\ . & . & . & \ldots & . \\ 0 & 0 & 0 & \ldots & 1 \\ 0 & 0 & 0 & \ldots & 0 \end{bmatrix}_{\alpha_i \times \alpha_i}, \quad b_i = \begin{bmatrix} 0 \\ 0 \\ \ldots \\ 0 \\ 1 \end{bmatrix}_{\alpha_i \times 1} \tag{14}$$

By specifying $z$ in Eq. (5) as

$$z = Ke + r \tag{15}$$

where $K = \mathrm{diag}(k_1, k_2, \ldots k_m)$ is the gain matrix and

$$\boldsymbol{k}_i = \begin{bmatrix} k_1^i & k_2^i & \cdots & k_{\alpha_i}^i \end{bmatrix}_{\alpha_i \times 1} \tag{16}$$

we may assign poles of the error dynamics system (13) to the desired position in the open left-half complex plane. Then $e = 0$ is an asymptotically stable equilibrium of the closed-loop system

$$\dot{e} = A_c e \tag{17}$$

where $A_c = A + BK$. Denote the exact solution of $u$ for Eq. (6) by $u_s$. Then $u_s$ achieves the asymptotically tracking to $y_r$. Now we will show that under certain conditions, the controller (12) realizes the asymptotically tracking as well.

**Assumption 1**: The exact solution of control for the implicit algebraic equation (6), i.e., $u_s$, exists, and the Jacobi matrix $\dfrac{\partial v}{\partial u}$ at the $u_s$ satisfies $\det(\dfrac{\partial v}{\partial u_s}) \neq 0$.

**Assumption 2**: The initial value $u_0$ in Eq. (12) is close to $u_s|_{t=0}$ such that $\det(\dfrac{\partial v}{\partial u}) \neq 0$.

**Lemma 1** [24]: For the linear time-invariant system (17), if system is asymptotically stable at the origin, then there exists a Lyapunov function

$$V_e = \frac{1}{2} e^T P e \tag{18}$$

which satisfies $\dot{V}_e(e) \leq 0$, where $P$ is a positive-definite matrix.

**Lemma 2** [24]: For the system (17) with the Hurwitz coefficient matrix $A_c$, the positive-definite matrix $P$ may be

$$P = 2\int_0^\infty e^{(A_c^T + A_c)t} dt \tag{19}$$

and we may derive that $\dot{V}_e \leq -e^T e$.

**Theorem 1**: For the system (1) and (2), if the ideal control $u_s$ achieves asymptotically tracking to the reference signal $y_r$, then under Assumptions 1 and 2, the controller (12) also asymptotically tracks $y_r$.

**Proof**: With the controller (12), the closed-loop error dynamics is

$$\dot{e} = A_c e + B h \tag{20}$$

where $h = v(x, u) - z$. Consider the Lyapunov function as

$$V_c = V_e + a V_s \tag{21}$$

where $V_s$ and $V_e$ are defined by Eqs. (7) and (18), respectively, $a = \dfrac{\max\left(\mathrm{eig}(B^T P^T P B)\right)}{2 \min\left(\mathrm{eig}(\dfrac{\partial v}{\partial u} K_s \left(\dfrac{\partial v}{\partial u}\right)^T)\right)}$ and eig() is the function of eigenvalue. Differentiating Eq. (21) gives

$$\dot{V}_c = -e^T e + e^T P B h - a h^T \frac{\partial v}{\partial u} K_s \left(\frac{\partial v}{\partial u}\right)^T h \tag{22}$$

According to the Young's inequality

$$e^T P B h \leq \frac{1}{2} e^T e + \frac{1}{2} h^T B^T P^T P B h \tag{23}$$

Then we have

$$\dot{V}_c \leq -\frac{1}{2} e^T e + h^T \left[\frac{1}{2} B^T P^T P B - a \frac{\partial v}{\partial u} K_s \left(\frac{\partial v}{\partial u}\right)^T\right] h \leq 0 \tag{24}$$

Thus, the error $e$ will converges to zero and this means the controller (12) asymptotically tracks the reference signal $y_r$. □

## 2.3 Further discussion

One may argue that by using a tip of taking the derivative of the original control, i.e., $\dot{u}$, as the new control inputs, the original system (1) is transformed to the one with affine form, and then the existing feedback linearization control method may be applied to derive the control law. This is true. In that way, the controller obtained is also of the integral form as

$$u = \int_{t_0}^{t} \left\{ -\left(\frac{\partial v}{\partial u}\right)^{-1} \left(\frac{\partial v}{\partial x} \dot{x} - z\right) \right\} dt, \ u(t_0) = u_0 \tag{25}$$

where $z$ is the correspondingly prescribed dynamics. However, this paper not only provides a different idea to achieve the feedback linearization controller design for general systems, but more importantly, the different controller form may bring extra convenience, as under some conditions we may apply an inverse-free version of Eq. (12) as

$$u = \int_{t_0}^{t} \left\{ -K_s \left(\frac{\partial v}{\partial u}\right)^{T} (v - z) \right\} dt, \ u(t_0) = u_0 \tag{26}$$

Consider the situation that the reference signal $y_r$ is a constant vector. Then the problem is simplified to be a set-point tracking. Since the set-point control problem can be reformulated as a regulation problem with simple coordinate translation, we consider the case of $y_r = \mathbf{0}$. Now the variables $e$ defined in Eq. (13) are part of the new coordinates. Moreover, with the Frobenius Theorem [1], we may find other coordinates $\eta$, which represent the zero dynamics, to form the diffeomorphism valid in a domain $\mathbb{D}$ containing the origin as

$$\begin{bmatrix} e \\ \eta \end{bmatrix} = T(x) \tag{27}$$

where $T(\cdot)$ represents the coordinate transformation and $T(\mathbf{0}) = \mathbf{0}$. Then the dynamics for the new coordinates are

$$\dot{e} = Ae + Bv \tag{28}$$

$$\dot{\eta} = f_0(\eta, e) \tag{29}$$

where $f_0$ determines the zero dynamics. Under the assumption that the zero dynamics $\eta$ is asymptotically stable, it will be shown that we may use controller (26) to stabilize the system.

**Assumption 3**: For the system $\dot{\eta} = f_0(\eta, \mathbf{0})$, the origin $\eta = \mathbf{0}$ is asymptotically stable in $\mathbb{D}$.

**Lemma 3** [16]: With the ideal feedback control $u_s = u_s(x)$, the system (1) and (2) is asymptotically stable under Assumption 3, and there is a continuously differentiable Lyapunov function $V_I$ such that in the domain $\mathbb{D}$

$$\dot{V}_I \leq -\Gamma(\|\eta\|_2) - d_1 \|e\|_2 \leq -d_2 \|\eta\|_2^2 - d_3 \|e\|_2^2 \tag{30}$$

where $\Gamma(\cdot)$ is a class $K$ function, and $d_i \ (i = 1, 2, 3)$ are some constants. Thus, $\lim_{t \to \infty} f(x, u_s) = \mathbf{0}$ and $\|f(x, u_s)\|_2^2 \leq L_1 \|x\|_2^2$ (or $\|f(x, u_s)\|_2^2 \leq L_2 \|e\|_2^2 + L_2 \|\eta\|_2^2$) holds in $\mathbb{D}$ for some constant $L_1$ (or $L_2$).

**Theorem 2**: For the system (1) and (2), if the ideal control $u_s$ achieves asymptotically stabilization to the constant reference signal $y_r$ in $\mathbb{D}$, then under Assumptions 1-3, the closed-loop system under controller (26) is also asymptotically stable with gain $K_s$ satisfying the matrix inequality

$$\frac{\partial v}{\partial u} K_s \left(\frac{\partial v}{\partial u}\right)^{T} - \frac{1}{2}(M + M^{T}) \geq \mathbf{0} \tag{31}$$

where $M = \left(\frac{\partial v}{\partial x} - \frac{\partial z}{\partial x}\right) \frac{\partial f}{\partial u} \bigg|_{u_s + \vartheta_1(u - u_s)} \left(\frac{\partial v}{\partial u}\right)^{-1} \bigg|_{u_s + \vartheta_2(u - u_s)}$, and $\vartheta_1$, $\vartheta_2$ are any constants within $[0, 1]$.

**Proof**: Consider the Lyapunov function candidate as

$$V_c = V_e + aV_s + bV_I \tag{32}$$

where $a$ and $b$ are some positive constants, and $V_s$, $V_e$ and $V_I$ are defined by Eqs. (7), (18) and (30), respectively. Differentiating Eq. (32) gives

$$\begin{aligned}\dot{V}_c &= -e^T e + e^T PBh - a\left(h^T \frac{\partial v}{\partial u} K_c \left(\frac{\partial v}{\partial u}\right)^T h + h^T \left(\frac{\partial v}{\partial x} - \frac{\partial z}{\partial x}\right) f(x,u)\right) + b\dot{V}_I \\ &= -e^T e + e^T PBh - a\left(h^T \frac{\partial v}{\partial u} K_c \left(\frac{\partial v}{\partial u}\right)^T h + h^T \left(\frac{\partial v}{\partial x} - \frac{\partial z}{\partial x}\right)(f(x,u) - f(x,u_s)) + h^T \left(\frac{\partial v}{\partial x} - \frac{\partial z}{\partial x}\right) f(x,u_s)\right) + b\dot{V}_I\end{aligned} \tag{33}$$

Using the mean value theorem we can obtain

$$\dot{V}_c = -e^T e + e^T PBh - a\left(h^T \frac{\partial v}{\partial u} K_c \left(\frac{\partial v}{\partial u}\right)^T h + \frac{1}{2} h^T (M + M^T) h + h^T \left(\frac{\partial v}{\partial x} - \frac{\partial z}{\partial x}\right) f(x,u_s)\right) + b\dot{V}_I \tag{34}$$

here $\vartheta_1$ and $\vartheta_2$ are some constants within $[0,1]$. According to the Young's inequality

$$e^T PBh \le e^T e + \frac{1}{4} h^T B^T P^T PBh \tag{35}$$

$$h^T \left(\frac{\partial v}{\partial x} - \frac{\partial z}{\partial x}\right) f \le \frac{1}{4a} h^T \left(\frac{\partial v}{\partial x} - \frac{\partial z}{\partial x}\right)\left(\frac{\partial v}{\partial x} - \frac{\partial z}{\partial x}\right)^T h + a f^T f \tag{36}$$

We have

$$\dot{V}_c \le -h^T \left[a \frac{\partial v}{\partial u} K_s \left(\frac{\partial v}{\partial u}\right)^T - \frac{a}{2}(M + M^T) - \frac{1}{4}\left(\frac{\partial v}{\partial x} - \frac{\partial z}{\partial x}\right)\left(\frac{\partial v}{\partial x} - \frac{\partial z}{\partial x}\right)^T - \frac{1}{4} B^T P^T PB\right] h + a^2 \|f(x,u_s)\|_2^2 + b\dot{V}_I \tag{37}$$

Use the results in Lemma 3 and set $b = \dfrac{2a^2 L_2}{\min(d_2, d_3)}$, we have

$$\dot{V}_c \le -h^T \left[a \frac{\partial v}{\partial u} K_s \left(\frac{\partial v}{\partial u}\right)^T - \frac{a}{2}(M + M^T) - \frac{1}{4}\left(\frac{\partial v}{\partial x} - \frac{\partial z}{\partial x}\right)\left(\frac{\partial v}{\partial x} - \frac{\partial z}{\partial x}\right)^T - \frac{1}{4} B^T P^T PB\right] h - a^2 L_2 e^T e - a^2 L_2 \eta^T \eta \tag{38}$$

To guarantee $\dot{V}_c \le 0$, we may set

$$\frac{\partial v}{\partial u} K_s \left(\frac{\partial v}{\partial u}\right)^T - \frac{1}{2}(M + M^T) - \frac{1}{4a}\left(\frac{\partial v}{\partial x} - \frac{\partial z}{\partial x}\right)\left(\frac{\partial v}{\partial x} - \frac{\partial z}{\partial x}\right)^T - \frac{1}{4a} B^T P^T PB \ge 0 \tag{39}$$

which is hold for any $\vartheta_1$ and $\vartheta_2$ within $[0,1]$. Since $a$ may be set arbitrarily large, thus from a limit viewpoint, we may cancel the last two terms to get the matrix inequality (31). □

For the matrix $K_s$ in Eq. (31), a feasible way to determine it may be setting $K_s = k_s I$, where $k_s$ is a positive constant and $I$ is a right dimensional identity matrix, and

$$k_s > \frac{\max\left(\text{eig}(M + M^T)\right)}{2\min\left(\text{eig}(\frac{\partial v}{\partial u}(\frac{\partial v}{\partial u})^T)\right)} \tag{40}$$

## 3 Aircraft outer-loop trajectory controller design

### 3.1 Mathematic model

The aircraft flight control includes the outer-loop guidance system and the inner-loop attitude control system. Here we focus on the design of the guidance system, i.e., the outer-loop trajectory controller, and the inner-loop controller same to the one in Ref. [7] is combined to realize the flight control. The aircraft trajectory motion may be described as [25]

$$\frac{dx}{dt} = V \cos\gamma \cos\chi \tag{41}$$

$$\frac{dy}{dt} = V\cos\gamma\sin\chi \tag{42}$$

$$\frac{dh}{dt} = V\sin\gamma \tag{43}$$

$$\frac{dV}{dt} = \frac{1}{m}[-D - mg\sin\gamma + T_x\cos\alpha\cos\beta + T_y\sin\beta + T_z\sin\alpha\cos\beta] \tag{44}$$

$$\frac{d\chi}{dt} = \frac{1}{mV\cos\gamma}\begin{bmatrix} L\sin\mu + Y\cos\mu + T_x(\sin\mu\sin\alpha - \cos\mu\cos\alpha\sin\beta) \\ +T_y\cos\mu\cos\beta - T_z(\cos\mu\sin\beta\sin\alpha + \sin\mu\cos\alpha) \end{bmatrix} \tag{45}$$

$$\frac{d\gamma}{dt} = \frac{1}{mV}\begin{bmatrix} L\cos\mu - Y\sin\mu - mg\cos\gamma + T_x(\sin\mu\sin\alpha - \cos\mu\cos\alpha\sin\beta) \\ -T_y\sin\mu\cos\beta + T_z(\sin\mu\sin\beta\sin\alpha - \cos\mu\cos\alpha) \end{bmatrix} \tag{46}$$

where $m$ is the mass of the aircraft, $x$, $y$ and $h$ represent the horizontal coordinate, the longitudinal coordinate and the height in the ground reference frame, respectively. $V$ is the velocity magnitude, $\gamma$ is the flight-path angle, and $\chi$ is the heading angle. $\alpha$ is the Angle-of-Attack (AoA). $\mu$ is the velocity back angle. $L = QSC_L$, $D = QSC_D$ and $Y = QSC_Y$ are the aerodynamic lift, drag, and side force, respectively. $C_L$, $C_D$ and $C_Y$ are the lift coefficient, drag coefficient, and side force coefficient, respectively. $S$ is the reference area. $Q$ is the dynamic pressure. $T_x$, $T_y$ and $T_z$ are the components of thrust $T$ along the aircraft body axis, and $g$ is the gravity acceleration. During the flight, the sideslip angle $\beta$ is regulated to be zero through the inner-loop attitude control system. Moreover, for $T_y$ and $T_z$ are small quantities, they are treated as disturbance to fascinate the controller design. Thus, Eqs. (44)-(46) are simplified as

$$\frac{dV}{dt} = \frac{1}{m}[-D - mg\sin\gamma + \eta T_{\max}\cos\alpha] \tag{47}$$

$$\frac{d\chi}{dt} = \frac{1}{mV\cos\gamma}[L\sin\mu + \eta T_{\max}\sin\mu\sin\alpha] \tag{48}$$

$$\frac{d\gamma}{dt} = \frac{1}{mV}[L\cos\mu - mg\cos\gamma + \eta T_{\max}\cos\mu\sin\alpha] \tag{49}$$

where $\eta = \dfrac{T}{T_{\max}}$ is the throttle and $T_{\max}$ is the maximum thrust.

### 3.2 Baseline controller

In the outer-loop trajectory control, the states are $\boldsymbol{x} = [x\ y\ h\ V\ \chi\ \gamma]^T$, and the control inputs are $\boldsymbol{u} = [\alpha\ \mu\ \eta]^T$. The outputs are the aircraft position coordinates $\boldsymbol{y} = [x\ y\ h]^T$ and the reference signal to be tracked is the position commands $\boldsymbol{y}_r = [x_r\ y_r\ h_r]^T$. According to the controller design procedure in last section, first we differentiate the outputs to give

$$\ddot{x} = \frac{\cos\chi\cos\gamma}{m}[-D + \eta T_{\max}\cos\alpha] - \left(\frac{\cos\mu\sin\gamma\cos\chi}{m} + \frac{\sin\mu\sin\chi}{m}\right)[L + \eta T_{\max}\sin\alpha] \tag{50}$$

$$\ddot{y} = \frac{\cos\gamma\sin\chi}{m}[-D + \eta T_{\max}\cos\alpha] - \left(\frac{\cos\mu\sin\gamma\sin\chi}{m} - \frac{\sin\mu\cos\chi}{m}\right)[L + \eta T_{\max}\sin\alpha] \tag{51}$$

$$\ddot{h} = \frac{\sin\gamma}{m}[-D + \eta T_{\max}\cos\alpha] + \frac{\cos\mu\cos\gamma}{m}[L + \eta T_{\max}\sin\alpha] - g \tag{52}$$

To achieve the desired tracking performance for $[x_r \ y_r \ h_r]^T$, we set

$$z = \begin{bmatrix} -k_{x1}(x-x_r) - k_{x2}(\dot{x}-\dot{x}_r) + \ddot{x}_r \\ -k_{y1}(y-y_r) - k_{y2}(\dot{y}-\dot{y}_r) + \ddot{y}_r \\ -k_{h1}(h-h_r) - k_{h2}(\dot{h}-\dot{h}_r) + \ddot{h}_r \end{bmatrix} \quad (53)$$

where $k_{i1}, k_{i2} \ (i = x, y, h)$ are gain parameters. Thus, we obtain the control algebraic equation as

$$v(x, u) - z = 0 \quad (54)$$

where

$$v = \begin{bmatrix} \dfrac{\cos\chi\cos\gamma}{m}[-D+\eta T_{max}\cos\alpha] - \left(\dfrac{\cos\mu\sin\gamma\cos\chi}{m} + \dfrac{\sin\mu\sin\chi}{m}\right)[L+\eta T_{max}\sin\alpha] \\ \dfrac{\cos\gamma\sin\chi}{m}[-D+\eta T_{max}\cos\alpha] - \left(\dfrac{\cos\mu\sin\gamma\sin\chi}{m} - \dfrac{\sin\mu\cos\chi}{m}\right)[L+\eta T_{max}\sin\alpha] \\ \dfrac{\sin\gamma}{m}[-D+\eta T_{max}\cos\alpha] + \dfrac{\cos\mu\cos\gamma}{m}[L+\eta T_{max}\sin\alpha] - g \end{bmatrix} \quad (55)$$

According to Eq. (12), we derive the outer-loop control law as

$$\begin{bmatrix} \alpha_c \\ \mu_c \\ \eta_c \end{bmatrix} = \int_{t_0}^{t} \left\{ -K \left(\dfrac{\partial v}{\partial u}\right)^T (v-z) - \left(\dfrac{\partial v}{\partial u}\right)^{-1} \left(\dfrac{\partial v}{\partial x}\dot{x} - \dot{z}\right) \right\} \mathrm{d}t, \quad \begin{bmatrix} \alpha_c \\ \mu_c \\ \eta_c \end{bmatrix}\bigg|_{t=t_0} = \begin{bmatrix} \alpha_0 \\ \mu_0 \\ \eta_0 \end{bmatrix} \quad (56)$$

Here the subscript "$c$" indicates the command. $\alpha_0$, $\mu_0$ and $\eta_0$ are the initial values of the control commands.

### 3.3 Robustness issue

In principle, the system dynamics must be perfectly known in the feedback linearization controller design. By far, we ignored the disturbance effect brought by the possible sideslip angle, the thrust components $T_y$ and $T_z$, and the aerodynamics model error. To increase the robustness of the controller, the integral control technique is employed to counteract those errors in maintaining the equilibrium flight [16], by adapting Eq. (53) as

$$z = \begin{bmatrix} -k_{x0}\int(x-x_r)\mathrm{d}t - k_{x1}(x-x_r) - k_{x2}(\dot{x}-\dot{x}_r) + \ddot{x}_r \\ -k_{y0}\int(y-y_r)\mathrm{d}t - k_{y1}(y-y_r) - k_{y2}(\dot{y}-\dot{y}_r) + \ddot{y}_r \\ -k_{h0}\int(h-h_r)\mathrm{d}t - k_{h1}(h-h_r) - k_{h2}(\dot{h}-\dot{h}_r) + \ddot{h}_r \end{bmatrix} \quad (57)$$

where $k_{i0} \ (i = x, y, h)$ are integral control gains.

### 4 Simulation Results

Based on a 6 degree-of-freedom nonlinear aircraft model, the aircraft level flight route maneuver example is simulated to examine the controller's performance. The initial position of the aircraft is $[x(t_0) \ y(t_0) \ h(t_0)]^T = [0 \ 0 \ 1000]^T$ m in the ground reference frame. The initial velocity is 50 m/s. The initial flight-path angle and heading angle are both 0 deg. The feedforward trajectory commands are $[x_r(t) \ y_r(t) \ h_r(t)]^T = [50(t-t_0) \ 50 \ 1050]^T$ and the flight time span is 25 s. In the control law (56), the initial values of control commands were set as $[\alpha_0 \ \mu_0 \ \eta_0]^T = [10\deg \ 0\deg \ 0.17]^T$. During the simulation, the thrust error and the aerodynamics error were considered, and the surface actuator and engine were modeled by first-order filters.

The aircraft position commands and the flight trajectory in 3-dimensional space are plotted in Fig. 1. It is shown that although the initial position is far from the set commands, the aircraft quickly changes its flight trajectory towards the expected route. Fig. 2 details the position tracking error profiles against time, it is found that the reference trajectory is well tracked and the position error is fairly small after 20 s. For comparison, the tracking error without integral control compensation is also presented, showing that the integral control effectively eliminate the steady-state error arising from the modeling errors.

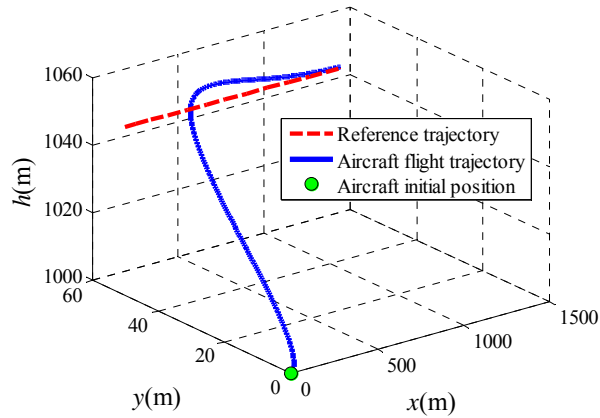

Fig 1. The 3-dimensional flight trajectory of the aircraft.

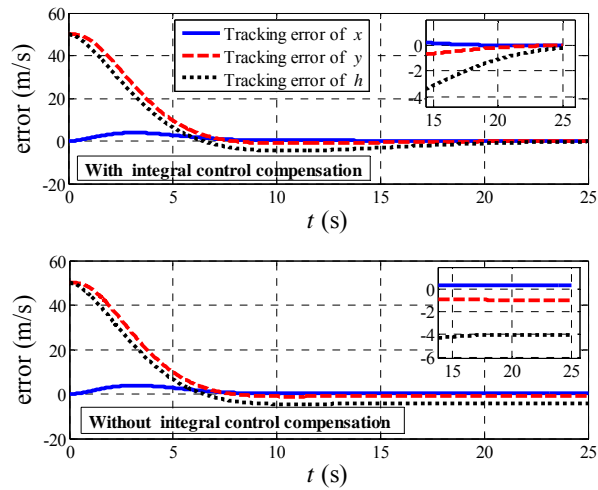

Fig 2. The position tracking error profiles with and without integral control

Fig. 3 gives the velocity magnitude profile, and Fig. 4 shows the heading angle and flight angle profiles of the aircraft. It is shown that after a sharp transition of 10s, the aircraft gradually realizes the level flight with the desired velocity of 50m/s.

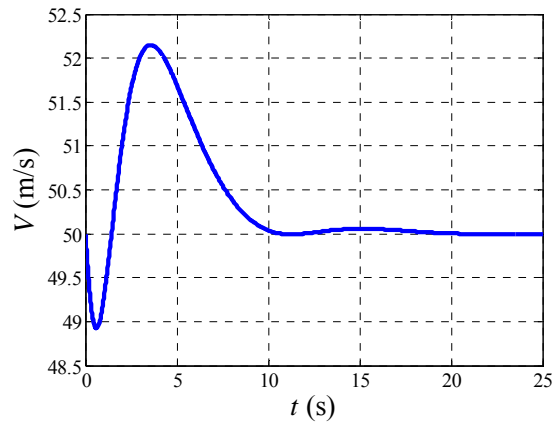

Fig 3. The velocity magnitude profile of the aircraft

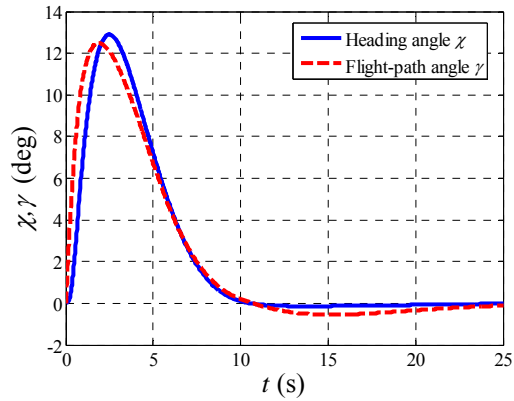

Fig 4. The heading angle and the flight-path angle profiles of the aircraft

Fig. 5 gives the AoA control commands and the velocity bank angle commands from the outer-loop trajectory controller. For the AoA control command, it sharply increases at first, and stays around the level flight angle after 10s. The velocity bank angle command also changes rapidly at the initial stage. Then it returns to vicinity of zero degree after 10s. The throttle command is presented in Fig. 6. It is shown that large thrust is demanded to change the trajectory initially, and then it gradually decreases to the trim value.

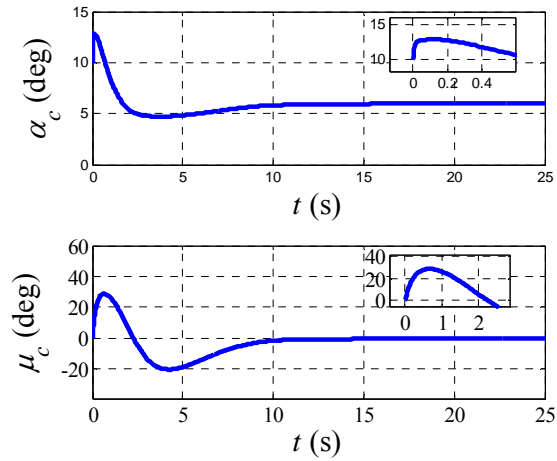

Fig 5. The AoA and the velocity bank angle commands from the trajectory controller

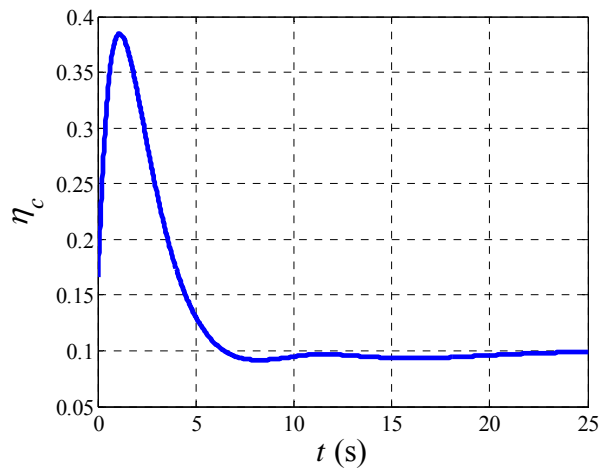

Fig. 6. The throttle command from the trajectory controller

In Fig. 7, the Lyapunov function defined in Eq. (7), which measures the error in solving the implicit control algebraic equation, is plotted. The error decreases to zero rapidly after 3s, which means the control commands is precise afterwards.

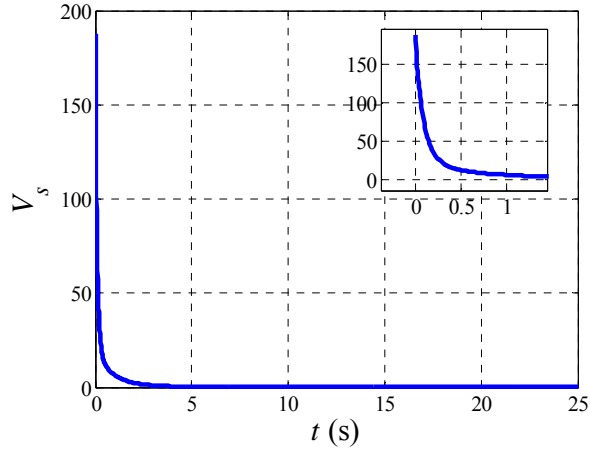

Fig 7. The error in solving the implicit control algebraic equations

Fig. 8 gives the aerodynamic control surface deflection curves including the aileron $\delta_a$, elevator $\delta_e$ and rudder $\delta_r$. They vary heavily in the first 5 seconds. Then all converge to deflection corresponding to the prescribed level flight.

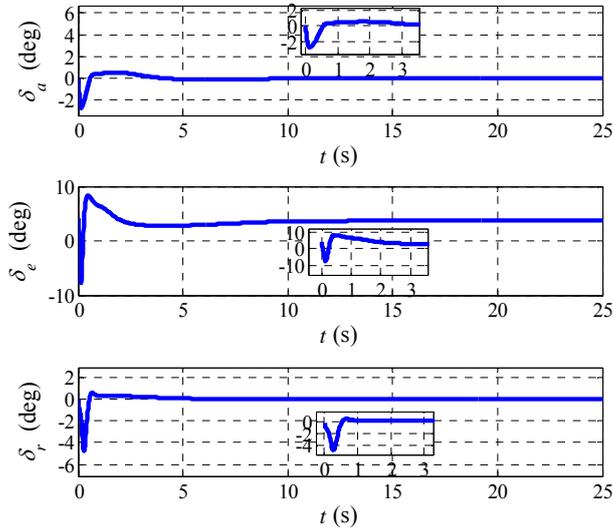

Fig 8. The aerodynamic control surfaces deflection in the flight

## 5　Conclusion

The feedback linearization control method is further developed for general systems to effectively address the implicit control algebraic equations encountered. The resulting control law is of appealing integral form upon augmented dynamics, and it is proved that the proposed controller may achieve asymptotically tracking. Moreover, theoretical analysis shows that the control law may be applied in an inverse-free form to the set-point control problem. This nonlinear control method is applied to the aircraft trajectory control, and the integral control technique is combined to address the robust concern. An aircraft level flight route maneuver (not limited to) example is simulated to verify the performance, and it is shown the developed controller works well to achieve the commanded trajectory.